# Columnar Database Techniques for Creating AI Features

Brad Carlile
AI Perf
Portland, OR
brad@aiperf.com

Akiko Marti
AI Perf
Portland, OR
akiko@aiperf.com

Guy Delamarter
AI Perf
Portland, OR
guy@aiperf.com

## ABSTRACT

Recent advances with in-memory columnar database techniques have increased the performance of analytical queries on very large databases and data warehouses. At the same time, advances in artificial intelligence (AI) algorithms have increased the ability to analyze data. We use the term AI to encompass both Deep Learning (DL or neural network) and Machine Learning (ML aka Big Data analytics). Our exploration of the AI full stack has led us to a cross-stack columnar database innovation that efficiently creates features for AI analytics. The innovation is to create Augmented Dictionary Values (ADVs) to add to existing columnar database dictionaries in order to increase the efficiency of featurization by minimizing data movement and data duplication. We show how various forms of featurization (feature selection, feature extraction, and feature creation) can be efficiently calculated in a columnar database. The full stack AI investigation has also led us to propose an integrated columnar database and AI architecture. This architecture has information flows and feedback loops to improve the whole analytics cycle during multiple iterations of extracting data from the data sources, featurization, and analysis.

## 1. INTRODUCTION

Recent years have witnessed an explosion of columnar databases [9] such as Oracle 12c Database In-Memory Option [10], AWS Redshift [6], MonetDB [7] and SAP HANA [5]. In-memory columnar databases provide additional advantages and can perform analytical queries 10x to 30x faster than row-format based storage.

In-memory columnar format databases make analytical queries faster, however transactional OLTP performance on the same data can suffer. Oracle 12c Database In-Memory Option, was the first dual-format database, allowing on-disk tables stored in row-format to also have a version of that data stored in-memory in columnar format. In the Oracle database both row-stored and column-stored data is maintained to be transactionally consistent during updates. This allows transactional processing and data warehousing to both operate on the latest data while providing efficient formats for both kinds of operations. Queries or transactions that are not optimal in row format can be dramatically faster in column format.

Hardware/software codesign has also created special hardware such as the SPARC DAX [1] [16] to provide a further 10x advantage in performance over software implementations of in-memory columnar databases [3].

Analytics on structured data stored in databases can be more than just SQL analytical queries. Databases are increasingly being used as inputs for machine learning (ML aka Big Data analytics), statistical processing, and deep learning (DL). We interchangeably use the term artificial intelligence (AI) to encompass both ML and DL.

We introduce innovations to improve in-memory columnar databases during the process of creating features for use with ML and DL analysis. These innovations use extensions to the dictionary encoding data structures used to compress data for columnar VLDBs. The efficiency of these innovations increases the value of using columnar databases for AI.

Given that, columnar databases are well suited for use with AI. We also explore the effects of this on the full-stack analytics pipeline. In our view the analytics pipeline should be viewed as an analytics cycle. It is very common to have multiple cycles of analysis. When viewed as an analytics cycle a richer set of data reuse, information flows and feedbacks between databases and AI are evident. In order to properly explain the innovations in this paper to the practitioners in the diverse areas of Database, ML, and DL, we will start with a survey of the key concepts involved.

## 2. THE SCOPE OF ANALYTICS

Traditional investigations into analytics have been narrow in scope. It is time to explore the full stack and look for optimizations and then design an appropriate overall architecture. Figure 1 shows the traditional scopes/silos of the various kinds of traditional analytics.

To database practitioners, the scope of analytics has focused on using analytical queries within OLTP and columnar databases to produce summaries and tables. In contrast, the approach many Big Data and Machine Learning (ML) practitioners have taken is to select data that is stored in file systems (CSV, JSON, …), then manipulate the data into the proper numerical format for subsequent ML algorithms (Linear Regression, Generalized Linear Models, Gradient Boosted Trees, SVD, …) to produce models and predictions. And finally DL practitioners have largely focused their analytics on image, video and text data using a variety of Deep Neural Networks (DNN) architectures to produce models and inferences. Research in all three of these areas has produced significant gains in performance, efficiency, and applicability.

We observe practitioners are beginning to move out of their traditional silos and the analytical approaches are merging. For instance, with the advent of technologies such as Amazon's RedShift and Amazon ML [2] now take a larger view and expand the traditional ML scope the complete way from columnar databases to feature engineering and finally to ML models.

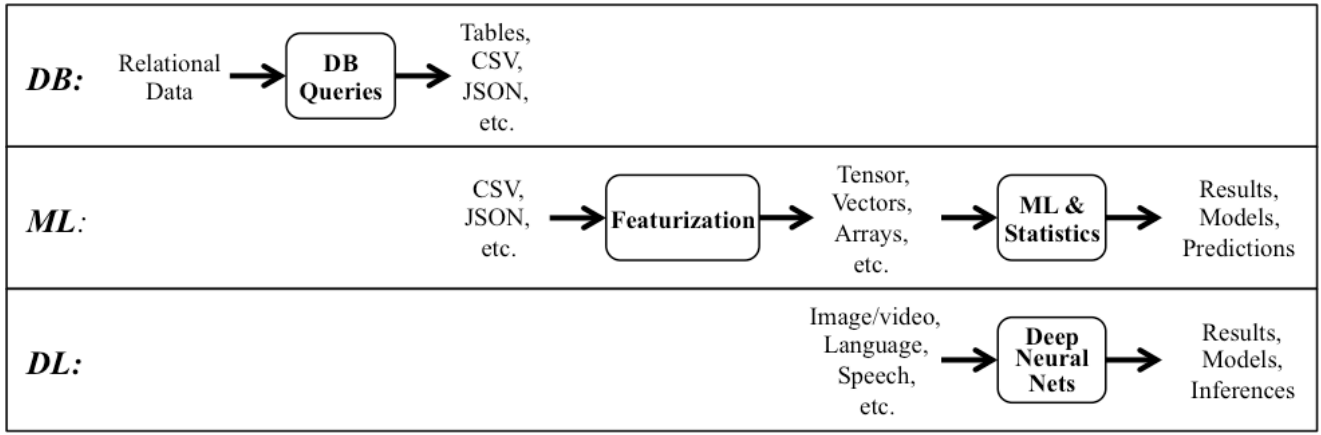

Figure 1. The traditional siloed view of the analytics pipeline. Full stack AI analytics will incorporate all of these areas.

In addition, technologies such as Google's Wide and Deep [4] and Microsoft's Wide and Deep Ensembles [12] are now combining generalized linear models (Wide) with deep learning (Deep) neural networks. Wide and Deep combines the strengths of both a wide linear model and a deep neural network. It has proven useful for generic large-scale regression and classification problems with sparse input features, such as those generated by OLTP transactions. We expect future research into techniques such as Wide and Deep to include other kinds of analysis and data types.

It is also interesting to note that many Big Data and ML practitioners are increasingly favoring SQL [3] as a powerful and concise language manipulating data while creating features for ML. SQL was developed as a powerful way to manipulate lots of data and to increase developer productivity by providing a high-level abstraction.

SQL has a rich legacy of well-known parallelization and optimization techniques for all modern databases. These optimizations are being applied to ML. In ML, SQL is sometimes leveraged indirectly from within other programming languages using in the form of a Domain Specific Language (DSL), or a Language-INtegrated Query (LINQ).

We are now approaching the time when we should be exploring performance optimizations across the entire analytics stack from DB to inference. In this paper we present innovations for the in-memory columnar for the purpose of improving the performance of feature engineering. The techniques described in this paper for improving columnar database SQL processing can be easily adapted for DSL use in Apache Spark and others analytics packages. The goal of these innovations is to advance DL and ML for full stack analytics.

In full stack AI, a wide variety of data and information will need to be accessed. The many traditional advantages of very large databases for securely managing data are well suited for AI analytics cycle. We need performance and efficiency at every level of the stack.

## 3. ANALYTICS CYCLE

The analytics process is traditionally called the analytics pipeline [12]. In the pipeline the data is selected, filtered, reduced dimensionally, and transformed (or "munged") to get it into the form required for ML or DL. The data is then analyzed with the appropriate algorithms and finally the model or the results generated.

It is rare that only one pass of the analytics is performed. Developing an understanding of subtleties of complex data generally requires multiple analyses or experiments. Often insights gained from one analysis are used to refine the next analysis. Thus we believe it is critical to recognize that the process is an analytical cycle. When viewed in this manner, various forms of data and analysis reuse are exposed. Optimizations to exploit reuse in the analytical cycle will clearly be valuable. We will explore this more in section 7.

## 4. DATA CHARACTERISTICS
### 4.1 Feature Characteristics

We next explore the characteristics of the data that will serve as input for analytics. Some of the important characteristics are the type of data, the range, and the uniqueness of its values. We can exploit the characteristics of data as we look for both efficient storage and eventual featurization.

In databases, we place data of the same type into each column (ex: age, gender, zip code, …). In ML each column can be a feature or the basis of a feature for analysis. Sometimes the data in a single ML feature is based on multiple columns.

There are two major classifications of features, categorical and numerical. Categorical features take on a fixed number of discrete values that have no particular mathematical ordering (ex: gender has no greater than or less than ordering). Typical categorical features are represented using strings or integers.

Numerical features are quantitative variables that can take on a range of numerical values and they have a mathematical ordering (ex: income). Many tend to divide numerical features into discrete or continuous classifications. Discrete features only have a small set of values (ex: number of living parents). Continuous features are generally considered to have a large number of possible values (ex: temperature). A more precise way to describe the number of possible values for any feature is cardinality.

## 4.2 Cardinality

Cardinality is the number of unique values a column (feature) may contain. For example, while one may have millions of US customers, they will reside in one of 50 states, so the state column would have a maximum cardinality of 50. Of course, the true cardinality of a particular dataset may be less than the maximum possible (ex: having only customers in 38 states).

In the process of data analysis we can use data bucketization (aka binning) to reduce the cardinality of the data. Bucketization will group high-cardinality data into a set of intervals. The number of buckets (cardinality) is often much less than the original.

Bucketization can be helpful in two ways, first it reduces the effects of minor observation errors and secondly it aligns the analysis with business decisions (ex: target marketing segments). Bucketization can be either linear or non-linear. An example of linear bucketization is to group people into equal size buckets (ex: 0-9, 10-19, 20-29, …). An example of a non-linear bucketization of "age life-cycle groups" (ex: 0-3, 4-12, 13-16, 16-21, 21-65,…).

## 5. COLUMNAR DATABASE

The advantages of columnar databases to accelerate performance of queries are well known [10]. By storing data in a columnar form, the database can access only the data it needs to perform a query without having to access the unneeded data in the rest of the row. Query performance is often dramatically faster. Many software and hardware techniques have been developed to accelerate performance on column databases such as compression.

## 5.1 Dictionary Columnar Compression

Compression can dramatically improve the performance of analytical queries by reducing the amount of data that needs to be read [11]. In addition, large multi-TB databases that could previously only be stored on disk can be completely stored in-memory if compression techniques are used. Examples of 2x to 30x compression have been achieved using multi-level compression techniques [9]. Columnar databases typically use dictionary compression and then this data additionally compressed with run-length encoded (RLE) or Huffman-based compression.

Dictionary encoding provides data compression and also captures metadata that are used for queries, featurization as well as ML and DL analytics. In dictionary encoding each original data value is mapped to an integer. For example, in a state column, the string "Alabama" is mapped to "1", "Alaska" is mapped to "2", and "The State of Rhode Island and Providence Plantations" is mapped to "39." Given that there are 50 states we can determine that only 6 bits are needed (ceil(log2(50))) in each column entry, which is a dramatic reduction in storage required. Columns of different bit-widths can be bit-packed in consecutive memory and quickly scanned using x86 SIMD instructions or the SPARC DAX co-processor. The dictionary encoding technique can also be used for analytics packages like Apache Spark SQL [3], which does processing on columnar data. Note that Apache Spark does not currently store columnar data in memory in a consecutive bit-packed format for fast scanning.

The mappings for the column entry encodings are stored in a dictionary. The dictionary may also store additional metadata to accelerate query processing such as column maximum and minimum. The Oracle Database [10] for example, uses the minimum and maximum metadata for optimizing queries with equality, in-list, and some range predicates. Notice that encoding values are internal to the database and may not have the same ordering as the data they are encoding.

**Table 1. Dictionary for state encoding for a column and maximum and minimum metadata for the column**

| Max = | 40 |
|---|---|
| Min = | 3 |
| **State** | **Encoding** |
| California | 3 |
| Connecticut | 7 |
| South Carolina | 40 |

Dictionary encoding often offers a particularly high degree of compression for string-type columns. This compression can be applied to both low-cardinality and high-cardinality data of all types. There can be significant compression depending on representation of the original data and the number of bits needed to encode it. Table 2 gives examples of feature data cardinalities for enterprise workflows.

**Table 2. Example cardinalities**

| Sample Features & Cardinalities | Characteristics | |
|---|---|---|
| | *Cardinality* | *Bits to Encode* |
| Binary Gender | 2 | 1.0 |
| Season | 4 | 2.0 |
| Marital Status | 5 | 2.3 |
| Months | 12 | 3.6 |
| US States | 50 | 5.6 |
| Age in Years | 150 | 7.2 |
| Countries in World | 195 | 7.6 |
| Day of Year | 366 | 8.5 |
| US Area Code | 999 | 10.0 |
| US Zip codes | 99,999 | 16.6 |
| Unique data (512k row) | 524,288 | 19.0 |

Oracle 12c Database In-Memory Option divides columns targeted for in-memory storage into In-Memory Compression Units (IMCUs). The typical size for an IMCU is half a million rows [10]. The maximum cardinality for an IMCU is 19-bits (2^19 = 512K). Dictionary compression is still advantageous for high cardinality (totally unique) data if it takes more than 19-bits per data element.

## 5.2 Additional Compression Methods

Dictionary encoding provides one level of compression. This data can be further compressed by applying other compression methods on top of dictionary-encoded data. Run-length encoding

(RLE) offers dramatic compression for sorted data and even semi-sorted data. RLE has also been found useful in practice on a wide variety of dictionary-encoded columnar data. Alternatively, Huffman compression or OZIP [10] compression can also be used.

# 6. COLUMNAR DB STRUCTURE FOR AI

As shown in Figure 1, the typical approach used by data scientists is to extract the data from databases and other sources and convert it to one or more files of an uncompressed universal format such as JSON or CSV. Then they use a variety of tools to explore the data and perform featurization. This approach inherently involves making one or more intermediate copies of the data, which is both inefficient and has security issues.

Since it is increasingly common for data scientists to use SQL or DSL-style SQL to both extract data from databases and transform them into features for ML frameworks, performing all of the whole operations in a database would be an easy transition.

With SQL one can already leverage the efficiencies of RDBMS and NoSQL databases manipulating large data at scale. Database execution benefits from a wide variety of innovations including: indexes, efficient joins, query optimizers, etc. In addition database optimizers have also become very efficient at extracting efficient parallelism from high-level SQL.

If the featurization is done close to where the data is stored then database innovations can be brought to bear on more of the process. There is less data movement, and fewer intermediate transformations (see Figure 1). Working within the database data can be transformed into exportable numerical features in a single efficient step. Columnar databases (RDBMS and NoSQL) are well suited to efficiently access just the data required to construct DL and ML features. Working entirely within the database for featurization the unsecure working copies of the data are avoided as well.

## 6.1 Feature Transformations for DL and ML

In this section we cover a variety of transformations that Data Scientists typically use when creating numerical inputs for DL and ML. We list some of the important methods used so database practitioners will be aware of the critical featurization operations that we will have the database perform. There are several different ways to get features ready for analysis. Feature selection is the process of selecting a subset of relevant features for model use. Feature creation (or feature transformation) is the process of creating a new feature from other information. For example, a feature for "single parent with single child" may be created from information on parents and number of total children.

### 6.1.1 Numeric type conversions

Data that is used in ML or AI must typically be in floating-point format as most algorithms are based on numerical analysis. Integer data will get converted to floating-point representation. 32-bit floating point require only 4 bytes to store in binary format, while in JSON or CSV it is up to 7 times larger since it can require up to 14 Unicode characters (28 bytes). This again points to the advantage of having databases perform featurization rather than having that process external.

### 6.1.2 Normalization

Most data exists at a wide variety of scales. It is best to standardize (or normalize) all data into similar ranges. Depending on the data there are a variety of manners to "standardize" data, these include: linear scaling, min/max scaling, mean normalization, standard deviation normalization, logarithmic scaling, z-score transforms, and others.

### 6.1.3 One-hot encoding

One-hot encoding maps a single low-cardinality column into multiple columns of binary vectors, each of which has at most a single one-value. Some also refer to one-hot encoding as "Categorical Identity Columns." This is an effective encoding for categorical data of low cardinality of both numerical and string types.

### 6.1.4 Binarizer, quantile, hash buckets, and bucketization

Binarizer, quantile, hash buckets, and bucketization discretization are all different types of cardinality reduction techniques for ML analysis.

Binarization is the process of thresholding numerical features to binary (either: 0.0 or 1.0) features. This can be done by linear thresholding or applying a logistic function. This can also be viewed as selective one-hot encoding.

Quantile discretization is the term used when the range of is divided into number of quantiles or bins. Typically with quantile discretization the range of input values is divided in a linear fashion. If the quantile discretization is set to create 4 buckets then the output would be 4 one-hot columns (a 4-element vector).

Hash buckets are another way to reduce cardinality of numeric columns. The hashing used is a modulo operator. While this may map completely unrelated features together it seems that this works well in practice, in part because features from other columns can be used to differentiate them.

Bucketization is the term used for dividing the feature into buckets where the ranges of each bucket may be different. Typically the range of input values is divided in a non-linear fashion. Sometimes the term bucketization is used as a general term to describe every type described in this section.

All of these discretizations serve two purposes in DL and ML. First, they produce lower cardinality representations which reduce the chances of over-fitting. Second, they can present data in a manner that is closer to the needs of the desired output. Does one really need to market to 42-year olds differently than 43-year olds, and if you should, could you?

### 6.1.5 Embedded features

High-cardinality data can be represented as short numerical vectors that are an embedding of the values into a smaller multidimensional space [13]. The vector representations themselves are typically learned during some analysis. Embedding can reveal relationships between features. Such embeddings have been very useful in natural language processing and have been applied to a wide variety of data types.

## 6.2 Count Metadata

Often queries return the count of various entries in a column. We suggest adding a count field to each dictionary entry. This would accelerate various kinds of queries without requiring any scan of the data, see Table 3.

**Table 3. Adding count to each dictionary entry**

| Max = | South Carolina | |
|---|---|---|
| Min = | California | |
| **State** | **Encoding** | **Count #** |
| California | 0 | 800 |
| Connecticut | 2 | 6 |
| South Carolina | 1 | 10 |

If the sum of a particular column is needed it can be calculated by using the partial sums directly from the dictionary without accessing the underlying data. Many other calculations such as average and standard deviation also require a sum. Feature normalization functions such as min/max scaling, mean normalization, standard deviation normalization can all be accelerated for many queries.

During the data exploration phase of analytics, a common operation is to create histograms. These and other feature transformations can be quickly computed with these count entries.

## 6.3 Augmented Dictionary Values (ADVs)

Many of the operations involved in featurization described in section 6.1 above can be accelerated by combining columnar dictionaries with an innovation we call Augmented Dictionary Values (ADVs).

For example a "State" column could get bucketized into 4 regions and could also get bucketized into 9 divisions. Since either may be useful in DL or ML we may want to maintain both mappings for the variety of analysis we wish to perform.

**Table 4. ADV entries added to dictionary with two bucketizations**

| Max = | California | | | |
|---|---|---|---|---|
| Min = | Virginia | | *ADVs* | |
| **State** | **Encoding** | **#** | **Region Bucket** | **Division Bucket** |
| California | 3 | 81 | 3.0 | 9.0 |
| Connecticut | 7 | 16 | 0.0 | 0.0 |
| Oregon | 6 | 37 | 3.0 | 8.0 |
| … | … | … | … | … |
| Virginia | 10 | 11 | 2.0 | 4.0 |

Since the output of these queries is for DL and ML processing which require floating-point format, the alternate feature mappings are populated with floating-point numbers of the type that can be directly used by the algorithms without conversion, see Table 4.

The ADVs used in Table 4 assume using several pre-designed buckets. Notice that bucketized ADVs store the index of the bucket and that for ML and DL these will actually be represented as multiple columns that are one-hot encoded.

Each ADV may store statistical measures of its data distribution in order to assess its interest as a feature, examples such as entropy, diversity, or peculiarity [15].

While a particular bucketization may align with business objectives or human intuitions, we also introduce the idea that new bucketizations could be learned during the course of analysis. These new bucketizations could be useful for future analytics in different workloads. We suggest it may be advantageous to present AI algorithms with multiple bucketizations and let the algorithm decide the best to use. Notice in Table 5 that one of the ADVs is Age in floating-point format, this is done to avoid integer to floating-point conversion at query time. As a reminder the encoding values are internal to the database and they are assigned to the various entries at load time. As in the example below they may have no correspondence to numerical column values.

**Table 5. Dictionary with a variety of ADVs bucketizations and one format conversion**

| Max = | 91 | | | | | | |
|---|---|---|---|---|---|---|---|
| Min = | 8 | | | *ADVs* | | | |
| **Age** | **Encoding** | **#** | **Decade Bucket** | **Age FP** | **Age Group** | **ML G1** | **DL G2** |
| 55 | 2 | 80 | 5.0 | 55.0 | 4.0 | 9.0 | 10.0 |
| 42 | 7 | 26 | 4.0 | 42.0 | 4.0 | 8.0 | 5.0 |
| 8 | 25 | 73 | 0.0 | 8.0 | 1.0 | 0.0 | 0.0 |
| … | … | … | … | … | … | … | … |
| 17 | 3 | 11 | 1.0 | 17.0 | 3.0 | 5.0 | 3.0 |

Columnar databases derive their efficiency on join and filter processing with dictionary encoding by performing simple calculations on small integers (1-bit to 32-bit) instead of their original format. SIMD instructions in x86 or purpose-built DAX hardware on the SPARC processor can operate very efficiently on these small integers. The dictionary is created once and then modified as the column is updated. Since there are many more analytical queries than updates in a data warehouse the overhead of creating and maintaining the dictionary is easily amortized. Only at the final stage of query processing, when the results are returned, does the original format need to be retrieved.

ML and DL featurization does the same filtering and joining as traditional analytical queries. The main differences are that different output formats are required. The simplest form of this format conversion needed is to provide a floating-point representation for ML and DL calculations. In a very general sense, the other featurization operations described in section 6.1 above can be viewed as various kinds of type conversions. Again if these are stored in a dictionary there is no need to calculate them, the can be looked up and outputted at the end of the query. Traditional ML featurization requires data to be read from a file, perform the join and filtering, and then calculate the output format. These extra operations are avoided by using ADVs and a columnar database.

The algorithms for updating ADVs during database inserts, updates, and deletes is a straightforward process. Obviously changes to the data may affect the results of any analysis.

## 7. COLUMMAR DATABASE AND AI ARCHITECTURE

We propose an optimized full-stack architecture encompassing a columnar database and AI analytics (DL and ML), to both maximize efficiency of data movement as well as address the needs of the analytics cycle, see Figure 2.

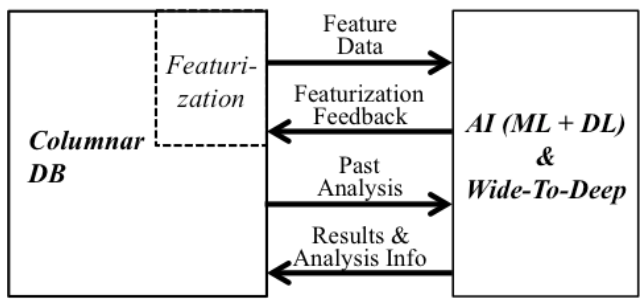

**Figure 2. Optimized columnar database and AI architecture.**

When we use the term columnar database we are referring both to Columnar RDBMS and Columnar NoSQL. The database should manage and store all data, metadata, featurization methods, and information around every analysis.

The guiding principles for this architecture are increasing data movement efficiency, providing performance at scale, and incorporating feedback based on learning to improve analysis. An example of the feedback mechanisms from the analysis to the columnar database is new bucketizations that are inferred from DL/ML analysis. These bucketizations may be useful for future analysis (see the generated bucketizations ML G1 and DL G2 in Table 5).

We expect data sources to contain a wide variety of data, metadata, hyperparameters, and etcetera about all aspects of AI. Examples of information stored and managed by the columnar database include, but are not limited to:

- created features, metadata on the features, and the calculations used to derive these created features;
- the analysis performed and the hyperparameters used for each analysis;
- weights and biases from previous deep learning neural nets to facilitate transfer learning;
- feedback on the importance/ranking/relevance of each feature, as well as, various feature bucketizations that worked well or were learned in the course of training;
- results from the analysis may also be used for future analysis.

Looking more closely at the various featurization operations we can see where new AI algorithms can provide feedback or improvement for subsequent analysis. First and foremost is the section of what type of featurization operation that should be performed. For example is it better to "Normalize" a column or to "Standardize" a column. Some of the operations below such as bucketization should only be applied to low-cardinality data for practical concerns about number of output columns. As shown below in Table 6, several of the featurization operations also have parameters. Various AI algorithms can also be created to provide improvement feedback for these operations.

**Table 6. Examples of Featurization grouped by similar functions**

| Featurization | Numerical (ordered) | Categorical (Unordered) | Parameters |
|---|---|---|---|
| Float | X | | |
| One-Hot Encoding | X | X | |
| Embedding | X | X | Embedding dimension |
| Min/Max | X | | |
| Normalization | X | | |
| Stardardization | X | | |
| Binarization | X | X (some operations) | Level / Cutoff |
| Quantile | X | | Number Quantiles |
| Bucketization | X | X | Bucketiz-ation Vector |

The database should also coordinate streaming data (ex: Kafka, logs, social feeds, IoT, ...) that are input to analysis. Samples from the stream or summary data of the data streams may also be appropriate for longer-term storage.

Databases also have the proper mechanism for collaborative work typical of most use cases in production analytics. Databases are also designed to have governance mechanisms for security and regulatory tracking.

In the future we expect Wide & Deep architectures to evolve to Wide-To-Deep (WTD) architectures that employ a range of techniques orchestrated by intelligent agents. We also expect multi-level architectures where for instance an ML result feeds into DL and ML to DL. These multi-level architectures may also both features that are passed through to the next level as well as have new features added at a level. In this evolution the value of the columnar database architecture and the ADV innovation will still hold.

## 8. CONCLUSION

The scope of the various disciplines of AI analytics is expanding to include every discipline in a unified full stack. In addition when we look at typical analytical workflows we find that the analytics pipeline is actually a rich analytic cycle that has data reuse as well as a variety of information flows and feedbacks. We also see many data types that can be analyzed with a variety of algorithms. Finally we see that all of the data and analytics need to be stored and managed in a secure fashion. This has led us to see the importance of an integrated database (data warehouse, NoSQL, RDBMs) combined with AI analytics.

It is well known that columnar databases are the most efficient for implementing data warehouses. The dictionary compression techniques used in columnar databases both reduce the data that needs to be stored and improve query execution. The ADV innovations described in this paper are additions to the dictionary for columnar data for accelerating the wide variety of featurization techniques that prepare data for AI analytics (ML & DL). The ADV columnar database innovation increases the efficiency of featurization by minimizing data movement and data duplication. We expect to seem more innovations for integrated columnar database and AI architecture as full stack AI is more fully explored.

## 9. ACKNOWLEDGMENTS

Our thanks to the other members of the AI Perf team, Brian Whitney and Paul Kinney, whose wise advice and reviews were greatly appreciated.